\documentclass[aps,prd,preprintnumbers,nofootinbib,floatfix,11pt]{revtex4}

\usepackage{amsfonts,amsmath,amssymb,mathrsfs,graphicx,epsfig,color,ulem,hyperref,ulem,xcolor}
\allowdisplaybreaks[3]
\newcommand{\be}{\begin{equation}}
\newcommand{\ee}{\end{equation}}
\newcommand{\ba}{\begin{eqnarray}}
\newcommand{\ea}{\end{eqnarray}}

\newcommand{\Cr}[1]{{{\cal O} \left( r^{#1} \right)}}

\newcommand{\blue}[1]{{\textcolor{black}{{#1}}}} 
\newcommand{\red}[1]{{\textcolor{black}{{#1}}}}

\DeclareRobustCommand{\erase}{\bgroup\markoverwith{\textcolor{blue}{\rule[.5ex]{2pt}{0.4pt}}}\ULon}
\DeclareRobustCommand{\Erase}{\bgroup\markoverwith{\textcolor{red}{\rule[.5ex]{2pt}{0.4pt}}}\ULon}

\begin{document}
\begin{flushright}
  YITP-21-103, KOBE-COSMO-21-15, OCU-PHYS-550, AP-GR-174
  \end{flushright}

\title{Asymptotic behavior of null geodesics near future null infinity II: curvatures, photon surface and dynamically transversely trapping surface}
\author{Masaya Amo$^1$, Tetsuya Shiromizu$^{2,3}$, Keisuke Izumi$^{3,2}$, Hirotaka Yoshino$^{4,5}$ and Yoshimune Tomikawa$^6$}

\affiliation{$^{1}$Center for Gravitational Physics, Yukawa Institute for Theoretical Physics, Kyoto University, 606-8502, Kyoto, Japan}
\affiliation{$^{2}$Department of Mathematics, Nagoya University, Nagoya 464-8602, Japan}
\affiliation{$^{3}$Kobayashi-Maskawa Institute, Nagoya University, Nagoya 464-8602, Japan} 
\affiliation{$^{4}$Advanced Mathematical Institute, Osaka City University, Osaka 558-8585, Japan}
\affiliation{$^{5}$Department of Physics, Kobe University, Kobe 657-8501, Japan}
\affiliation{$^{6}$Faculty of Economics, Matsuyama University, Matsuyama 790-8578, Japan}

\begin{abstract}
Bearing in mind our previous study on asymptotic behavior of null geodesics near future null infinity, 
we analyze the behavior of geometrical quantities such as a certain extrinsic curvature and Riemann tensor 
in the Bondi coordinates. In the sense of asymptotics, the condition
for an $r$-constant hypersurface to be a photon surface 
is shown to be controlled by 
a key quantity that determines the fate of photons initially emitted in angular directions. 
In four dimensions, such a non-expanding photon surface can be  
realized even near future null infinity
in the presence of enormous energy flux for a short period of time. 
By contrast, in higher-dimensional cases, no such a 
photon surface can exist.
This result also implies that the dynamically transversely trapping
surface, which is proposed as an extension of a photon surface,
can have an arbitrarily large radius in four dimensions.
\end{abstract}

\maketitle

%
%
%
\section{Introduction}

In recent years, there have been unprecedented reports of various observations on black holes: 
the gravitational-wave observations (see \cite{Abbott:2016blz}
for the first detection) and the shadow imaging \cite{Akiyama:2019cqa}. 
In the discussion of systems far away from massive objects,
as in black hole observations, 
ideal observers are considered to stay at future null infinity 
in asymptotically flat spacetimes, which are formulated precisely in 
Refs. \cite{Bondi,Sachs,Penrose:1962ij,Penrose:1965am,Ashtekar:1978, Tanabe:2009va,Tanabe:2011es, Hollands:2003ie,Hollands:2003xp}. 
Thus\blue{,} it is important to understand the properties of the asymptotic structure near future null infinity. 
Although the spacetime asymptotes to the Minkowski spacetime near infinity, 
there are non-trivial features of the asymptotics. 
One of them is the
supertranslation, which gives an infinite number of independent generators for symmetries \cite{Bondi,Sachs}. 
By contrast with the cases in four dimensions, however,  supertranslation\blue{s} are absent in higher dimensions if 
one supposes the finiteness of global charges such as mass \cite{Tanabe:2011es, no-memory}. Recently, supertranslation\blue{s} attract much attention in several 
topics including the memory effect \cite{Zeldovich:1974,Christodoulou:1991,Thorne:1992} 
and the soft theorem \cite{Strominger:2014pwa}. 

In our previous paper \cite{Amo:2021gcn}, meanwhile, the authors examined null geodesics that correspond to worldlines 
of photons emitted in angular directions of the Bondi coordinates near future null infinity. 
Surprisingly, there exists 
a non-trivial difference between four and higher dimensions (See Ref.~\cite{Cao:2021mwx} for the extension 
for Brans-Dicke theory). 
In higher dimensions, any of these 
null geodesics always reaches future null infinity. 
In four dimensions, by contrast, it is not guaranteed: 
gravitational waves and the flow of matter energy could affect
the fate of the null geodesics 
(See Sec.~\ref{review} for a brief review and/or Ref.~\cite{Amo:2021gcn} for detail).

The behavior of geodesics is imprinted in the geometrical quantities, namely curvatures. 
One of the examples is the photon sphere, which describes the unstable circular orbits of photons in the Schwarzschild black hole spacetime. 
That is to say, although the photon sphere indicates a collection of specific null geodesics,
equivalent conditions can be given
with the geometrical quantities as well~\cite{Virbhadra:1999nm, Claudel:2000}.
The concept of the photon sphere has been extended to general spacetimes \cite{Claudel:2000}, called the photon surface\blue{, which is the collection of certain (not necessarily circular) photon orbits}.
The generalization of the photon surface has been widely discussed
\cite{Shiromizu:2017ego,Yoshino:2017gqv,Siino:2019vxh,Yoshino:2019dty,Cao:2019vlu} \blue{(See also Ref. \cite{Koga:2019uqd} for the stability for the photon surface)}. 
In this paper, 
such an existing idea for photon spheres (or photon surfaces) is applied to the cases for the 
asymptotic behavior of null geodesics analyzed in our previous work \cite{Amo:2021gcn}, that is, 
we shall show the properties of the asymptotic null geodesics emitted in the angular direction near future null infinity 
in terms of the extrinsic curvature and the Riemann tensor. 
Our previous work \cite{Amo:2021gcn} suggests that 
there is an essential difference between the cases of four and higher dimensions. 
We will see similar differences in the geometrical quantities, 
{\it i.e.} their 
nontrivial asymptotic features can be seen only in four-dimensional cases. 

It is known that a hyperboloid in the Minkowski spacetime
is a photon surface, 
and thus, an arbitrarily large expanding photon surface
can be introduced in general \cite{Claudel:2000}.
In this paper, in contrast to such an expanding photon surface,
we focus on the condition for an $r$-constant hypersurface
to be a photon surface (say, a non-expanding photon surface).
The formation of such a photon surface
is fairly nontrivial, implying the existence of strong gravity. 
We will see in four dimensions that non-expanding photon surfaces can exist (locally) in the asymptotic region 
if enormous outgoing energy flux is present. 
In addition, we will show that 
the non-expanding photon surface described above 
is, at the same time, 
the dynamically transversely
trapping surface (DTTS) proposed by four of us
in order to describe the strong gravity region
in terms of behavior of photons \cite{Yoshino:2019dty}.
This indicates that a DTTS with an arbitrarily large radius 
can form near future null infinity. 
Most of these studies will be done in an approximate way by 
adopting the leading-order terms in the powers of $1/r$, but an exact analysis 
using the Vaidya spacetime 
will also be briefly reported. 

The rest of this paper is organized as follows. In Sec.~\ref{review}, we give a brief review of asymptotically 
flat spacetimes in terms of the Bondi coordinates and our previous work \cite{Amo:2021gcn}. 
In Sec.~\ref{psf}, we show the extrinsic curvature of the surface
given by a constant radial coordinate near future null 
infinity and discuss the presence of an approximate photon surface.
In Sec.~\ref{Sec:DTTS}, we see the fact that the result of Sec.~\ref{psf}
implies that an arbitrarily large DTTS can form near future null infinity.
In Sec.~\ref{curv}, we show asymptotic 
behavior of the Riemann tensor. 
Section~\ref{Conclusions} is devoted to a summary and discussion.
In Appendix~\ref{Appendix:Vaidya}, we present the calculations for a Vaidya spacetime.

%

\section{Asymptotic behavior of spacetime and null geodesics} 
\label{review}

In this section, we briefly review the spacetime behavior near future null infinity following 
Refs.~\cite{Bondi,Sachs,Tanabe:2011es} (see also \cite{Hollands:2003ie,Hollands:2003xp,Ishibashi:2007kb}) 
and our previous results on null geodesics \cite{Amo:2021gcn}. 

Let $n (\geq 4)$ be the dimension of spacetimes. 
For asymptotically flat spacetimes, the metric near future null infinity is written in the Bondi coordinates as
\begin{eqnarray}
    \label{Bondi}
ds^2 =g_{\mu\nu}dx^\mu dx^\nu= -Ae^B du^2 -2 e^B du dr + h_{IJ}r^2(dx^I + C^I du)(dx^J + C^J du),
\end{eqnarray}
where the Greek indices denote the spacetime components. 
$A, B, C^I$ and $h_{IJ}$ are functions of $u$, $r$ and $x^I$. Here, $x^I$ stands for the angular coordinates. 
Future null infinity is supposed to be in the limit of $r \to \infty$ while $u$ is finite. 
We impose the gauge condition as  
\begin{align}
    \label{gau}
    \sqrt{\det h_{IJ}}=\omega_{n-2},
\end{align}
where $\omega_{n-2}$ is the volume element of the unit $(n-2)$-dimensional sphere $S^{n-2}$. 
The functions $A$, $B$, $C^I$ and $h_{IJ}$ can be expanded
with respect to power of $1/r$, whose explicit formulas
are shown in Ref.~\cite{Tanabe:2011es}. 
The vacuum Einstein equation shows us that the nonzero components of the metric behave as \footnote{In even dimensions, 
each exponent in the Landau symbol of Eq.~\eqref{metric} actually has the higher order by $r^{1/2}$, but we write in 
the same way as in odd dimensions for unification. 
Note that quantities below, such as the Christoffel symbols, do not have half-integer powers with respect to $1/r$ 
in even dimensions.}
\begin{eqnarray}
g_{uu}&=&-Ae^B+h_{IJ}C^IC^J r^2 =-1-A^{(1)}r^{-(n/2-1)}+mr^{-(n-3)} +\Cr{-(n-1)/2}, \nonumber\\
g_{ur} &=& -e^B =-1 -B^{(1)}r^{-(n-2)} + \Cr{-(n-3/2)},\nonumber\\ 
g_{IJ} &=& h_{IJ}r^2 = \omega_{IJ}r^2 + h^{(1)}_{IJ}r^{-(n/2-3)} + \Cr{-(n-5)/2}, \nonumber \\
\quad g_{uI}&=& h_{IJ} C^J r^2 = C^{(1)}_{~~I}r^{-(n/2-2)} +\Cr{-(n-3)/2},\label{metric}
\end{eqnarray}
where $\omega_{IJ}$ denotes the metric of the unit $(n-2)$-dimensional sphere. 
\blue{In general relativity, the integration of $m(u,x^I)$ over the solid angle gives the Bondi mass, 
\begin{eqnarray}
M(u):=\frac{n-2}{16\pi}\int_{S^{n-2}}md\Omega \label{M(u)}.
\end{eqnarray}}
We can also apply the fall-off behavior of Eq. \eqref{metric} to non-vacuum spacetimes if the fall-off behavior of the stress-energy tensor is sufficiently fast such that the lowest order of the Einstein tensor behaves as $G_{\mu\nu}=\Cr{-n/2}$ in the coordinate system which asymptotes to the Cartesian coordinate system near null infinity. For example, the stress-energy tensor of the Maxwell field in even dimensions behave as $T_{\mu\nu}=\Cr{-(n-2)}$ in the same coordinate, and then the current setup \red{works}. The example in Appendix A also satisfies the fall-off behavior of the stress-energy tensor for the $n=4$ case. 
We list some components of the inverse metric and the Christoffel symbols which we will use later: 
\ba
g^{rr}&=&1+A^{(1)}r^{-(n/2-1)}-mr^{-(n-3)}+\Cr{-(n-1)/2}\label{g^rr}
\ea
and 
\ba
    \Gamma^{r}_{uu}
    &=& \frac{1}{2}\dot{A}^{(1)}r^{-(n/2-1)}-\frac{1}{2}\dot{m}r^{-(n-3)} + \Cr{- (n-1)/2}, \quad 
    \nonumber \\
    \Gamma^{r}_{ur}
    &=&- \frac{n-2}{4}A^{(1)}r^{-n/2}+\frac{n-3}{2}mr^{-(n-2)}+\Cr{-(n+1)/2},\nonumber\\
     \Gamma^{r}_{rr}&=& 
      -(n-2)B^{(1)}r^{-(n-1)}+\Cr{-(n-1/2)}, \quad 
    \nonumber \\
    \Gamma^{r}_{uI}
    &=& \left(\frac{n-4}{4}C^{(1)}_I+\frac{1}{2}A^{(1)}_{,I}\right)r^{-(n/2-1)}+\left(-\frac{1}{2}m_{,I}+\frac{1}{2}C^{(1)J}\dot{h}^{(1)}_{IJ}\right)r^{-(n-3)}  \nonumber \\
    &&\hspace{6mm}+\Cr{-(n-1)/2} ,\nonumber\\
     \Gamma^{r}_{rI} &=& 
      \frac{n}{4}C^{(1)}_Ir^{-(n/2-1)}+\Cr{-(n-1)/2}, \quad 
    \nonumber \\
    \Gamma^{r}_{IJ}
    &=& -\omega_{IJ} r+\frac{1}{2}\dot{h}^{(1)}_{IJ}r^{-(n/2-3)} +\Cr{-(n-5)/2}, \label{Christoffel}
\ea
where $C^{(1)}_I:=C^{(1)J}\omega_{IJ}$, and $A^{(1)}$ is set to be zero for $n = 4$ because it is absorbed into $m$. 
The variables with dots, such as $\dot A^{(1)}$, denote their derivatives with respect to $u$. 

We will look at the asymptotic behavior of null geodesics near future null infinity (see our previous paper 
\cite{Amo:2021gcn} for the details). Let us focus on null geodesics that correspond to worldlines of photons 
emitted in the tangential directions to the $r$-constant surfaces near future null infinity, {\it i.e.} the ones with  
$r^\prime=0$, where $^\prime$ denotes the derivative with respect to the affine parameter. At the emission point, 
the $r$-component of the 
geodesic equation is calculated as
\ba
r''\blue{|_{r'=0}} &=&-\Gamma^r_{\mu\nu}\left(x^\mu\right)^\prime\left(x^\nu\right)^\prime \nonumber \\
    &=& \Big[ \omega_{IJ} r-\frac{1}{2}\dot{h}^{(1)}_{IJ}r^{-(n/2-3)}+\frac12\dot{m}\omega_{IJ}r^{-(n-5)}-\frac{1}{2}\dot{A}^{(1)}\omega_{IJ}r^{-(n/2-3)} \nonumber\\
&&\hspace{12mm} +\Cr{-(n-5)/2}\Big] \left(x^I\right)^\prime \left(x^J\right)^\prime,\label{r''_withoutu'}
\ea
by using the null condition $g_{\mu\nu}\left(x^\mu\right)^\prime\left(x^\nu\right)^\prime=0$, the future directed condition $u^\prime\geq0$, and Eq.~\eqref{Christoffel}. 
In four dimensions, Eq.~\eqref{r''_withoutu'} becomes
\ba
r''\blue{|_{r'=0}} &=&\left[\Omega_{IJ}r+\Cr{0}\right]\left(x^I\right)^\prime \left(x^J\right)^\prime\label{r''_final_4d},
\ea
where $\Omega_{IJ}$ is defined as
\ba
\Omega_{IJ}&:=&\omega_{IJ}-\frac{1}{2}\dot{h}^{(1)}_{IJ}+\frac12\dot{m}\omega_{IJ}.\label{Omega}
\ea
Therefore, even at sufficiently large $r$, 
the trajectory of a photon is not approximated by that in the flat spacetime, but determined by $\Omega_{IJ}$.  
Moreover, the sign of $r^{\prime\prime}$ is determined by that of the eigenvalues of $\Omega_{IJ}$, 
and thus, $r^{\prime\prime}$ can be negative. In Ref.~\cite{Amo:2021gcn},
we have proved that
the null geodesics will 
 reach future null infinity provided that $\Omega_{IJ}$ is positive definite and $\dot{m}\leq0$ in four dimensions. 
Unless these conditions are satisfied, 
the null geodesics with the same initial conditions may not be able to reach 
future null infinity. 
One may expect that 
$\dot h^{(1)}_{IJ}$ and $\dot{m}$ would be sufficiently small 
and $\Omega_{IJ}$ would be positive definite for almost all situations.
However, the possibility that photons do not reach
future null infinity would be fairly surprising.

In the case $n\geq5$, Eq.~\eqref{r''_withoutu'} becomes
\ba
r''\blue{|_{r'=0}} &=&\left[\omega_{IJ}r+\Cr{-(n/2-3)}\right]\left(x^I\right)^\prime \left(x^J\right)^\prime. 
\label{r''_final_nd}
\ea
This implies that 
the trajectory of photons is approximated by that in the flat spacetime and
any null geodesic always reaches future null infinity \cite{Amo:2021gcn}.

%
%
%

\section{
Extrinsic curvature
and photon surface}
\label{psf}

The behavior of geodesics depends on the geometry of a spacetime. 
This gives us an expectation that the properties corresponding to the non-trivial asymptotic behavior of null geodesics 
described in the previous section can be seen in the geometric quantities. 
In the discussion of the photon sphere (or the photon surface), such a relation 
has been shown \cite{Claudel:2000}.  
Photon surfaces are defined to be nowhere-spacelike codimension-one hypersurface $S$ such that, for every point 
$p \in S$ and every null vector $k^a\in T_p S$ (small latin indices run the coordinate $u$ and angular coordinate 
indices \blue{$I$}), a null geodesic $\gamma$ 
tangent to $k^a$ at $p$ is included in $S$\blue{ at least for a finite section of the geodesics around $p$. }
Alternatively, an equivalent condition of the photon surface is given in terms of the geometric quantities, {\it i.e.},
$S$ is a photon surface if and only if $\chi_{ab}k^a k^b=0$ holds for all null tangent vectors $k^a$ to $S$, where $\chi_{ab}$ 
is the extrinsic curvature. This condition is also equivalent to the condition that the surface is umbilical, {\it i.e.} 
vanishing of the traceless part of the extrinsic curvature, $\sigma_{ab}$, of $S$, {\it i.e.} $\sigma_{ab}=0$ \cite{Claudel:2000,Perlick:2005jn}. 
Similarly, the non-trivial behavior of asymptotic null geodesics found in four dimensions would be reflected in
the asymptotic behavior of the geometric quantities, and
it is natural to expect that the $r$-constant hypersurface with vanishing $\Omega_{IJ}$ defined by Eq.~\eqref{Omega} becomes 
a photon surface. 

We begin this section with the calculation of the extrinsic curvature $\chi_{ab}$ of the $r$-constant hypersurface 
$S_r$ with sufficiently large $r$. The induced metric on $S_r$ is given by 
\ba
P_{ab}dx^adx^b=-Ae^Bdu^2+h_{IJ}r^2(dx^I+C^Idu)(dx^J+C^Jdu),
\ea
that is, each component of the metric is written in 
\ba
P_{uu}&=&-Ae^B+h_{IJ}C^IC^J r^2 =-1-A^{(1)}r^{-(n/2-1)}+mr^{-(n-3)} +\Cr{-(n-1)/2}, \label{P_uu}\\
\quad P_{uI}&=& h_{IJ} C^J r^2 = C^{(1)}_Ir^{-(n/2-2)} +\Cr{-(n-3)/2}, \label{P_ui} \\
P_{IJ} &=& h_{IJ}r^2 = \omega_{IJ}r^2 + h^{(1)}_{IJ}r^{-(n/2-3)} + \Cr{-(n-5)/2}.\label{P_IJ}
\ea
For later convenience, we also write down the inverse of $P_{ab}$,
\ba
P^{uu}&=&-A^{-1}e^{-B}=-1+A^{(1)}r^{-(n/2-1)}-mr^{-(n-3)}+\Cr{-(n-1)/2},\\
P^{uI}&=&A^{-1}e^{-B}C^I=C^{(1)I}r^{-n/2}+\Cr{-(n+1)/2},\\
P^{IJ}&=&h^{IJ}r^{-2}-A^{-1}e^{-B}C^IC^J=
\left[\omega^{IJ}-h^{(1)IJ}r^{-(n/2-1)}+\Cr{-(n-1)/2}\right]r^{-2}.
\ea

The extrinsic curvature of the $r$-constant hypersurface is 
\ba
\chi_{ab} := \nabla_a r_b = -\frac{1}{\sqrt{g^{rr}}}\Gamma^r_{ab}\label{chi},
\ea
where we used the fact that the outward unit normal vector \blue{$r_a$ is written as 
$r_a=\left(g^{rr}\right)^{-1/2}(dr)_a$} in the Bondi coordinates. Using Eqs. \eqref{g^rr} and \eqref{Christoffel},  
the components of $\chi_{ab}$ are calculated as 
\ba
\chi_{uu} &\blue{=}&-\blue{\left[1+\Cr{-1}\right]}\Gamma^r_{uu}= -\frac12\dot{A}^{(1)}r^{-(n/2-1)}+\frac12\dot{m}r^{-(n-3)}+\Cr{-(n-1)/2},\label{chi_uu}\\
\chi_{uI} &\blue{=}&-\blue{\left[1+\Cr{-1}\right]}\Gamma^r_{uI} = -\left(\frac{n-4}{4}C^{(1)}_I+\frac12A^{(1)}_{,I}\right)r^{-(n/2-1)}\nonumber\\
&&\hspace{45mm}+\left(\frac12m_{,I}-\frac12C^{(1)J}\dot{h}^{(1)}_{IJ}\right)r^{-(n-3)}+\Cr{-(n-1)/2},\\
\chi_{IJ} &\blue{=}&-\blue{\left[1+\Cr{-1}\right]}\Gamma^r_{IJ}=\omega_{IJ}r-\frac12\dot{h}^{(1)}_{IJ}r^{-(n/2-3)}+\Cr{-(n-5)/2}\label{chi_IJ}.
\ea
In the current setup, it is easy to see 
\ba
r''\blue{|_{r'=0}} =-\Gamma^r_{\mu\nu}k^\mu k^\nu\blue{=\left[1+\Cr{-1}\right]}\chi_{ab}k^ak^b \label{Gamma-chi}
\ea
on $S_r$. 
We can see the relation between the extrinsic curvature $\chi_{ab}$ and the behavior of null geodesic momentarily tangent to 
$S_r$. Equation \eqref{r''_final_4d} shows that, in four dimensions, 
the behavior of null geodesics cannot be approximated by that in the flat spacetime
because of the existence of $\dot{m}$ and $\dot{h}^{(1)}_{IJ}$ in $\Omega_{IJ}$. 
We can see in Eqs.~\eqref{chi}--\eqref{chi_IJ} that the non-trivial asymptotic properties are 
imprinted in the extrinsic curvature $\chi_{ab}$ in four dimensions: in the leading order of $\chi_{ab}k^a k^b$, $\dot{m}$ 
and $\dot{h}^{(1)}_{IJ}$ appear together
with $\chi_{ab}k^a k^b$ of the flat spacetime.
We have seen, moreover, that $r''$ becomes negative if and only if $\Omega_{IJ} k^I k^J$ is negative, 
and then $r$ decreases at least for a short period of time. 
Here, $k^I$ is composed of the angular components of a null vector $k^a$ tangent to $S_r$. 
Equations \eqref{r''_final_4d} and \eqref{Gamma-chi} show that the condition for the 
negativity of $\Omega_{IJ} k^I k^J$ corresponds to that of $\chi_{ab}k^a k^b$. 
By contrast, in higher dimensions, $\chi_{ab}k^a k^b$ is always positive, 
which is consistent with the fact that any null geodesics always reach future null infinity. 
The leading contributions are the same as those in the flat spacetime, 
and thus the extrinsic curvature can be approximated by that of the flat spacetime.

Let us see the condition where the  $r$-constant hypersurface becomes a photon surface. 
It is equivalent to the traceless part of the extrinsic curvature vanishes ($\sigma_{ab}=0$) 
or $\chi_{ab}k^a k^b=0$ for any  null vector $k^a$ tangent to $S_r$. 
The traceless part of the extrinsic curvature is written in
\ba
\sigma_{ab}:=\chi_{ab}-\frac{1}{n-1}\chi P_{ab},\label{sigma}
\ea
where $\chi$ denotes the trace part of $\chi_{ab}$. 
Each components of $\sigma_{ab}$ is 
evaluated as 
\ba
\sigma_{uu} &=& \frac{n-2}{n-1}r^{-1}-\frac{n-2}{2(n-1)}\dot{A}^{(1)}r^{-(n/2-1)}+\frac{n-2}{2(n-1)}\dot{m}r^{-(n-3)}+\Cr{-(n-1)/2},\\
\sigma_{uI} &=& -\left(\frac{n^2-n-4}{4(n-1)}C^{(1)}_I+\frac12A^{(1)}_{,I}\right)r^{-(n/2-1)}\nonumber\\
&&\hspace{20mm}+\left(\frac12m_{,I}-\frac12C^{(1)J}\dot{h}^{(1)}_{IJ}-\frac{1}{2(n-1)}\dot{A}^{(1)}C^{(1)}_I\right)r^{-(n-3)}\nonumber\\
&&\hspace{20mm}+\frac{1}{2(n-1)}\dot{m}C^{(1)}_Ir^{-(3n/2-5)}+\Cr{-(n-1)/2},\\
\sigma_{IJ} &=& \frac{1}{n-1}\omega_{IJ}r-\frac12\dot{h}^{(1)}_{IJ}r^{-(n/2-3)}-\frac{1}{2(n-1)}\dot{A}^{(1)}\omega_{IJ}r^{-(n/2-3)}+\frac{1}{2(n-1)}\dot{m}\omega_{IJ}r^{-(n-5)}\nonumber\\
&&\hspace{20mm}+\Cr{-(n-5)/2}.
\ea
\blue{Motivated by \blue{the} property of the photon surface that $\sigma_{ab}=0$, we call \blue{a retarded time interval $u_i\red{<} u \red{<} u_f$ of the $r$-constant hypersurface}
$S_r$ near future null infinity an ``approximate photon surface" when $\sigma_{ab}=0$ at the leading order in $r^{-1}$ expansion \blue{on this retarded time interval of $S_r$}.}
One can see that the leading-order contributions of $\sigma_{ab}$ are generically non-zero,  
which means that an approximate photon surface does not exist in general near future null infinity. 
This is quite reasonable. 
However,  only in four dimensions, 
there is a possibility that the leading order contributions  are canceled with each other, and
the results in the previous section suggest that the photon surface would 
appear if 
$\Omega_{IJ}$ vanishes. 

In four dimensions, $\sigma_{ab}$ becomes  
\ba
\sigma_{uu}&=&\left(\frac{2}{3}+\frac{1}{3}\dot{m}\right)r^{-1}+\Cr{-2}\label{sigma_uu_4d},\\
\sigma_{uI}&=&\sigma_{Iu}=\Cr{-1},\\
\sigma_{IJ}&=&\left(\frac13\omega_{IJ}-\frac12\dot{h}^{(1)}_{IJ}+\frac16\dot{m}\omega_{IJ}\right)r+\Cr{0}\label{sigma_IJ_4d}.
\ea
The condition for the approximate photon surface, {\it i.e.} for $\sigma_{ab}=0$ at the leading order, is that
\ba
\dot m=-2 ~~~ \mathrm{and} ~~~ \dot{h}^{(1)}_{IJ}=0\label{asy_psf_condition}
\ea
are satisfied \footnote{In the normalized basis, $\sigma_{\hat u \hat u}=\Cr{-1}$, $\sigma_{\hat u \hat I}=\Cr{-2}$ and 
$\sigma_{\hat I \hat J}=\Cr{-1}$, where the hatted indices denote the components with respect to the normalized basis. 
  Since the order of $\sigma_{\hat u \hat I}$ is higher compared the others,
  there is no requirement from $\sigma_{uI}$.}. 
\blue{Note that these conditions are not satisfied simultaneously for vacuum spacetimes in general relativity.} \blue{$\dot m=-2$ describes the existence of outgoing matter flux, which determines the rate of change of the Bondi mass. $\dot{h}^{(1)}_{IJ}=0$ describes the absence of gravitational wave radiation.}
From the definition of $\Omega_{IJ}\blue{:=\omega_{IJ}-\frac{1}{2}\dot{h}^{(1)}_{IJ}+\frac12\dot{m}\omega_{IJ}}$, we see that it is equivalent to 
\ba
\Omega_{IJ}=0. \label{Omegazero}
\ea
\blue{Here, we used $\omega^{IJ}h^{(1)}_{IJ}=0$, which is a consequence of the gauge condition Eq. \eqref{gau}.}
As seen in the previous section, it indicates that $r''$ vanishes if $r'=0$, 
that is, $r$ does not change and thus geodesics stay on $S_r$.

By contrast, in dimensions higher than four, the traceless part of the extrinsic curvature $\sigma_{ab}$ becomes
\ba
\sigma_{uu} &=& \frac{n-2}{n-1}r^{-1}+\Cr{-(n/2-1)},\\
\sigma_{uI} &=& \sigma_{Iu}=\Cr{-(n/2-1)},\\
 \sigma_{IJ} &=& \frac{1}{n-1}\omega_{IJ}r+\Cr{-(n/2-3)}.
\ea
All of the leading terms of $\sigma_{ab}$ do not vanish 
and they are the same as the components of $\sigma_{ab}$ in the flat spacetime.
Thus, there is no 
approximate photon surface for spacetimes whose dimensions are higher than four.

Note that, since the above study only took account of the leading order terms,
the photon surface is an approximate one: the $r$-constant
surface satisfies
the condition for the photon surface with the error of $\mathcal{O}(1/r)$.
As discussed in Appendix~\ref{Appendix:Vaidya}, 
for an outgoing Vaidya solution, which is an exact solution
representing the spherically symmetric with the null outgoing matter,
it is possible to make a situation
where an $r$-constant hypersurface becomes a photon surface exactly.
This requires a fine tuning of the behavior of the Bondi mass $M(u)=m(u)/2$.

We emphasize again that the difference of the behavior of the null geodesics near null 
infinity between four dimensions and higher dimensions 
arises even when $|\dot{m}|$ and $\dot{h}^{(1)}_{IJ}$ are not so large as $\dot{m}\sim-2$.
The trajectory is affected even by small $\dot{m}$ and $\dot{h}^{(1)}_{IJ}$ in four dimensions, 
but is not in higher dimensions. \blue{This is similar to the memory effect in the sense that both of them lead to critical differences between four dimensions and higher dimensions due to the asymptotic behavior.}
We have shown here that this effect is understood in terms of the extrinsic curvature.

%

\section{DTTS formation near future null infinity}
\label{Sec:DTTS}

We have seen in Sec.~\ref{psf} that in a four-dimensional spacetime,
the behavior of null geodesics can be drastically 
different from that in the Minkowski spacetime 
due to the difference of the extrinsic curvature, and 
the $r$-constant hypersurface becomes an approximate photon surface
if the conditions of Eq.~\eqref{asy_psf_condition} are satisfied.
This  immediately implies that the formation
of a dynamically transversely trapping surface (DTTS)
near future null infinity is possible as well in four dimensions.
The DTTS has been defined as an extension of a photon surface
by four of us in Ref.~\cite{Yoshino:2019dty}.
The DTTS has an analogy with an apparent horizon,
and thus, it is calculable on a spacelike hypersurface 
in generic spacetimes.
To be specific, an $(n-2)$-dimensional closed spacelike surface
$\sigma_0$ is called a DTTS if there is an $(n-1)$-dimensional timelike surface $S$ that contains $\sigma_0$ and 
satisfies the following three conditions on $\sigma_0$:
\begin{eqnarray}
  \bar{k} &=& 0, \label{non-expanding}\\
  \mathrm{max}\left(\bar{K}_{ab}k^ak^b\right) &=& 0,\label{marginally-transversely-trapping}\\
         {}^{(3)}\bar{\pounds}_{\bar{n}}\bar{k} & \le & 0.
         \label{accelerated_contraction}
\end{eqnarray}
Here, we span the time coordinate $t$
starting from $\sigma_0$ in $S$ so that its lapse function
is constant on each $t$-constant surface, and 
$\bar{k}$ denotes the trace of the extrinsic curvature 
of $t$-constant surfaces in $S$. $\bar{K}_{ab}$ is the extrinsic curvature
of $S$, and $k^a$ is an arbitrary null vector tangent to $S$.
${}^{(3)}\bar{\pounds}_{\bar{n}}$ denotes the
Lie derivative with respect to the unit normal $\bar{n}$ to $\sigma_0$
in $S$. 
If the equality in the inequality of Eq.~\eqref{accelerated_contraction}
holds, $\sigma_0$ is called a marginal\blue{ly} DTTS. 

The physical meaning of this definition is as follows. From $\sigma_0$
we emit photons in tangential directions to $S$.
The condition of Eq.~\eqref{non-expanding} means that
$S$ is chosen
so that $\sigma_0$ is an extremal surface in $S$.
The condition of Eq.~\eqref{marginally-transversely-trapping}
determines how the surface $S$ bends in the neighborhood of $\sigma_0$:
the emitted photons must propagate inside of $S$ or on $S$, and
for each point on $\sigma_0$, at least one photon must propagate on $S$.
Then, we consider the time slice of $S$, and
the condition of Eq.~\eqref{accelerated_contraction}
implies that $\sigma_0$ is a DTTS if $\sigma_0$ is a maximal
surface.

\blue{We call} \blue{a section of a $u$-constant surface $S_{u,r}$ in $S_r$} \blue{near future null infinity an ``approximate marginally DTTS" when Eqs.~\eqref{non-expanding} \eqref{marginally-transversely-trapping}, and the equality in the inequality of Eq.~\eqref{accelerated_contraction} are satisfied at the leading order in $r^{-1}$ expansion.}
Note that the DTTS, $\sigma_0$, is supposed to be a closed surface in the definition to exclude a trivial one such as a plane in flat spacetime.
But we do not explicitly impose this condition for the approximate DTTS because by ignoring higher-order contributions, $S_{u,r}$ is almost closed with a coarse-grained sense although it may not be necessarily closed. 
For a spherically symmetric spacetime discussed in Appendix A, one can construct a closed one.

Let us confirm that the $u$-constant
surface $S_{u,r}$ in the approximate photon surface $S_r$ constructed
above in the four-dimensional case is \blue{an approximate marginally DTTS} \footnote{
  In Ref.~\cite{Shiromizu:2017ego},
  four of us defined a loosely trapped surface (LTS), 
which is another concept to characterize a strong gravity region.
The LTS is defined as an $(n-2)$-dimensional
surface on some spacelike hypersurface whose mean curvature $k$ satisfies
$dk/dr \geq 0$. 
If we introduce the spacelike hypersurface by $t=\mathrm{constant}$ with $t=u+r$\blue{ in the Bondi coordinates}, 
the mean curvature $k$ of the $r$-constant surface
is evaluated as $k \simeq 
2/r-m/r^2$ 
and $dk/dr\simeq (-2+\dot{m})/r^2$ cannot be non-negative.}.
Since the photon surface $S_r$ satisfies $\bar{K}_{ab}k^ak^b=0$
(here $\bar{K}_{ab}=\chi_{ab}$),
the $u$-constant surface satisfies the condition
of Eq.~\eqref{marginally-transversely-trapping}.
For the family of $u$-constant hypersurfaces,
the trace $\bar{k}$ of the extrinsic curvature in $S_r$ is
\begin{equation}
\bar{k} = {}^{(3)}\bar{\pounds}_{\bar{n}} \log \sqrt{\mathrm{det}\left(r^2h_{IJ}\right)} = \Cr{-2} \label{bark}
\end{equation}
from the gauge condition of Eq.~\eqref{gau}. Here we used 
$\bar n=(-P^{uu})^{-1/2}\partial_u-P^{Iu}(-P^{uu})^{1/2}\partial_I \simeq 
[1+\Cr{-1}]\partial_u-C^{(1)I}r^{-2}\partial_I$. Then, the condition of Eq.~\eqref{non-expanding} 
is approximately satisfied. 
Although the lapse function of the
$u$ coordinate on $S_r$ is not constant on each $u$-constant
surface in general, 
the change in its value is $\mathcal{O}(1/r)$,
and the condition of Eq.~\eqref{accelerated_contraction}
is also approximately satisfied. Therefore,
the section of $u$-constant and $S_r$ is \blue{an approximate marginally DTTS}.

The above approximate study can be performed exactly in the spherically symmetric
Vaidya spacetime. Due to the spherical symmetry,
the lapse function of the coordinate $u$ is constant on each 
$u$-constant hypersurface. 
In this case, Eq.~\eqref{bark} becomes $\bar{k}=0$.
From the calculation of Appendix~\ref{Appendix:Vaidya},
the section of an $r$-constant hypersurface
and a $u$-constant hypersurface
becomes a DTTS if
\begin{equation}
  \frac{dM}{du} \le -\left(1-\frac{2M}{r}\right)\left(1-\frac{3M}{r}\right)
  \label{DTTS_condition_Vaidya}
\end{equation}
taking account of
all orders with respect to the powers of $1/r$, where $M$ is the Bondi mass defined in Eq.~\eqref{M(u)}.
This means that in four dimensions, an arbitrarily large DTTS can form. 
Although in Ref.~\cite{Yoshino:2019dty} four of us have proved the Penrose-like inequality
\begin{equation}
A\le 4\pi (3GM_{0})^2
\end{equation}
for the area $A$ of any DTTS on a time-symmetric spacelike hypersurface with a certain condition such as negativity 
of the radial pressure, 
where $M_{0}$ is the Arnowitt-Deser-Misner (ADM) mass, 
this areal inequality does not hold in general.
Enormously large energy flux, whose radial pressure is positive, can create a strong gravity field
even near future null infinity in the sense of the effect on 
the motion of transversely emitted photons.

%

\section{Riemann Curvature of 
$S_r$}
\label{curv}

In this section, we shall examine the inner geometry of the $r$-constant surface $S_{r}$ 
looking at the Riemann tensor for sufficiently large $r$. The calculations are done for general situation, {\it i.e.}, in general dimensions with $\dot{h}_{IJ}^{(1)}$ being generically nonzero.
The case where $S_r$ becomes an approximate photon surface in four dimensions
is briefly commented.

After short calculations, we have the $(n-1)$-dimensional 
Riemann tensor $^{(n-1)}R^{a}_{~bcd}$ as
\ba
^{(n-1)}R^{u}_{~IuJ}
&\simeq&\partial_u^{~(n-1)}\Gamma^u_{~IJ}\nonumber\\
&=&\frac{1}{2}\ddot{h}^{(1)}_{IJ}r^{-(n/2-3)}+\Cr{-{(n-5)/2}},\\
^{(n-1)}R^{u}_{~IJK}
&\simeq&D_J^{~(n-1)}\Gamma^u_{~IK}-D_K^{~(n-1)}\Gamma^u_{~IJ}\nonumber\\
&=&\frac{1}{2}\left(D_J\dot{h}^{(1)}_{IK}-D_K\dot{h}^{(1)}_{IJ}\right)r^{-(n/2-3)}+\Cr{-(n-5)/2},\\
^{(n-1)}R^{I}_{~JKL}
&=&^{(\omega)}R^I_{~JKL}+\frac{1}{4}\left(\dot{h}^{(1)I}_{~~~K}\dot{h}^{(1)}_{JL}-\dot{h}^{(1)I}_{~~~L}\dot{h}^{(1)}_{JK}\right)r^{-(n-4)}+\Cr{-(n/2-1)},\label{Rie_n-1_ang}
\ea
where $^{(\omega)}R^I_{~JKL}$ is the Riemann tensor of the $(n-2)$-dimensional round sphere, that is, 
$^{(\omega)}R_{IJKL}=\omega_{IK}\omega_{LJ}-\omega_{IL}\omega_{KJ}$. 
In the derivation of the above, the Christoffel symbol for the induced metric $P_{ab}$ is required to be calculated, 
\ba
^{(n-1)}\Gamma^u_{uu} 
&\simeq& \frac{1}{2}P^{uu}\partial_uP_{uu} \nonumber\\
&=& \frac{1}{2}\dot{A}^{(1)}r^{-(n/2-1)}-\frac{1}{2} \dot{m}r^{-(n-3)}+\Cr{-(n-1)/2},\\
^{(n-1)}\Gamma^u_{uI} 
&\simeq& \frac{1}{2}P^{uu}\partial_I P_{uu}+\frac{1}{2}P^{uJ}\partial_u P_{IJ} \nonumber\\
&=& \frac{1}{2}A^{(1)}_{,I} r^{-(n/2-1)}-\frac{1}{2}m_{,I}r^{-(n-3)}+\frac{1}{2}C^{(1)J}\dot{h}^{(1)}_{IJ}r^{-(n-3)}+\Cr{-(n-1)/2},\\
^{(n-1)}\Gamma^u_{IJ} 
&\simeq& -\frac{1}{2}P^{uu}\partial_u P_{IJ}\nonumber\\
&=& \frac{1}{2}\dot{h}^{(1)}_{IJ}r^{-(n/2-3)}+\Cr{-{(n-5)/2}},\\
^{(n-1)}\Gamma^I_{uu} 
&\simeq& P^{IJ}\partial_uP_{Ju}\nonumber\\
&=& \dot{C}^{(1)I}r^{-n/2} +\Cr{-(n+1)/2} ,\\
^{(n-1)}\Gamma^I_{uJ} 
&\simeq& \frac{1}{2}P^{IK}\partial_u P_{KJ} \nonumber\\
&=& \frac{1}{2}\dot{h}^{(1)I}_{~~~J}r^{-(n/2-1)}+\Cr{-(n-1)/2},\\
^{(n-1)}\Gamma^I_{JK} 
&\simeq& \frac{1}{2}P^{IL}\left(\partial_J P_{LK}+\partial_K P_{LJ}-\partial_L P_{JK}\right) -\frac12P^{Iu}\partial_uP_{JK}\nonumber\\
&=& ^{(\omega)}\Gamma^I_{JK}+\Gamma^{(1)I}_{~~~JK}r^{-(n/2-1)}- \frac12 C^{(1)I} \dot h^{(1)}_{JK} r^{-(n-3)} +\Cr{-(n-1)/2},
\ea
where $^{(\omega)}\Gamma^I_{JK}$ denotes the Christoffel symbol with respect to $\omega_{IJ}$, $\Gamma^{(1)I}_{~~~JK}$ is defined as
\ba
\Gamma^{(1)I}_{~~~JK}:=\frac{1}{2}\left(D_J h^{(1)I}_{~~~K}+D_K h^{(1)I}_{~~~J}-D^I h^{(1)}_{IK}\right),
\ea
$D_I$ is the covariant derivative with respect to $\omega_{IJ}$, and $D^I$ denotes $\omega^{IJ}D_J$. 

In the normalized basis, we see that the orders of the first two components of the Riemann tensor become 
$^{(n-1)}R^{\hat u}_{~\hat I \hat u \hat J}=\Cr{-(n/2-1)}$ and $^{(n-1)}R^{\hat u}_{~\hat I \hat J \hat K}=\Cr{-n/2}$, where 
the hatted indices denote the components with respect to the normalized basis.
For the third component,
we have $^{(n-1)}R^{\hat I}_{~\hat J \hat K \hat L}=\Cr{-2}+\Cr{-(n-2)}$.

For four dimensions, Eq.~\eqref{Rie_n-1_ang} becomes
\ba
^{(3)}R^{I}_{~JKL}&=&^{(\omega)}R^I_{~JKL}+\frac{1}{4}\left(\dot{h}^{(1)I}_{~~~K}\dot{h}^{(1)}_{JL}-\dot{h}^{(1)I}_{~~~L}\dot{h}^{(1)}_{JK}\right)+\Cr{-1}. \label{3Riemann}
\ea
This is a quite impressive result because, as seen soon, the term such as the second one in the right-hand side of 
Eq.~\eqref{3Riemann} does not appear in dimensions higher than four. The reason of the appearance in four dimensions
is that the Ricci scalar of a round sphere with the radius $r$ 
is $\mathcal{O}(1/r^2)$, while the amplitude of gravitational waves 
decay as $\mathcal{O}(1/r)$. This behavior is related 
to the memory effect originated from supertranslation\blue{s}. 

One can show that the leading-order term of $^{(3)}R^{I}_{~JKL}$ is identical to the Riemann tensor of the round sphere 
if and only if $\dot h_{IJ}^{(1)}=0$. Under the condition $\dot{h}^{(1)}_{IJ}=0$, it is trivial to 
show that the leading term of $^{(3)}R^{I}_{~JKL}$ coincides with the Riemann tensor of the round sphere. 
Conversely, when the leading term of $^{(3)}R^{I}_{~JKL}$ is identical to the Riemann tensor of the round sphere at the 
leading order, 
\ba
\dot{h}^{(1)}_{IK}\dot{h}^{(1)}_{JL}-\dot{h}^{(1)}_{IL}\dot{h}^{(1)}_{JK}&=&0\label{spherical_condition}
\ea
holds. With Eq.~\eqref{gau} and four-dimensional speciality, it is easy to see that Eq.~\eqref{spherical_condition} implies $\dot{h}_{IJ}^{(1)}=0$. 
\blue{See Ref. \cite{Geroch} for similar discussion.}
To sum up, $\dot{h}^{(1)}_{IJ}=0$ is the necessary and sufficient condition for the leading part of 
$^{(3)}R^I_{~JKL}$ to be identical to the Riemann tensor of the round sphere. 
In addition, we can see 
that the other components of the Riemann tensor fall off as ${}^{(3)}R^{\hat u}_{~\hat I \hat u \hat J}=\Cr{-1}$ and ${}^{(3)}R^{\hat u}_{~\hat I \hat J \hat K}=\Cr{-2}$. 
Here, we recall the case where $S_{r} $ is the photon surface at the leading order
in four dimensions examined in Sec.~\ref{psf}. 
In that case, $\Omega_{IJ}$ vanishes, which gives $\dot{h}^{(1)}_{IJ}=0$ on $S_{r} $. Thus, 
one can see that $^{(3)}R^I_{~JKL}$ is identical to the Riemann tensor of the round sphere at the leading order. 

We now discuss the higher-dimensional cases. In these cases, 
Eq.~\eqref{Rie_n-1_ang} becomes
\ba
^{(n-1)}R^{I}_{~JKL}&=&^{(\omega)}R^I_{~JKL}+\Cr{-(n-4)}+\Cr{-(n/2-1)},\label{Rie_n-1_ang2}
\ea
and thus, the leading term of $^{(n-1)}R^{I}_{~JKL}$ is identical to the Riemann tensor of the sphere.
The other components of $^{(n-1)}R^{a}_{~bcd}$ rapidly decay for $n\ge7$. 
For five dimensions, $^{(4)}R^{\hat u}_{~\hat I \hat u \hat J}=\Cr{-3/2}$ is larger than 
$^{(4)}R^{\hat I}_{~\hat J \hat K \hat L}=\Cr{-2}$, and, for six dimensions, 
$^{(5)}R^{\hat u}_{~\hat I \hat u \hat J}=\Cr{-2}$ is comparable to
$^{(5)}R^{\hat I}_{~\hat J \hat K \hat L}=\Cr{-2}$.

It is merely trivial from the setup that, at the leading order in both four dimensions and higher dimensions, the Riemann tensor ${\cal R}^{I}_{~JKL}$ with 
respect to the induced metric on the $u$-constant surface in $S_r$ is identical to 
that of the round sphere. 
It is easily obtained as 
\ba
{\cal R}^{I}_{~JKL}&\simeq& ^{(\omega)}R^I_{~JKL}+\left(D_K \Gamma^{(1)I}_{~~~JL}-D_L \Gamma^{(1)I}_{~~~JK}   \right)r^{-(n/2-1)}.
\ea

%
\section{Summary and Discussion}
\label{Conclusions}

In this paper, adopting the Bondi coordinates, we have analyzed the asymptotic behavior of the extrinsic curvature 
and the Riemann tensor
of the $r$-constant surface 
near future null infinity. In particular, we explored the relations to our previous study, 
that is, asymptotic behavior of null geodesics. Therein, in four dimensions, one could see that the tensor 
$\Omega_{IJ}$ defined by Eq.~\eqref{Omega} determines the fate of photons emitted in angular directions. 
As a consequence,  
the non-trivial properties of asymptotic behavior of null geodesics
have been shown in terms of 
the geometric quantities, and
we have confirmed the direct relation 
between vanishing of $\Omega_{IJ}$ and the umbilical feature 
of the non-expanding photon surface. 
This result also implies that an arbitrarily large DTTS
can be formed if enormously large energy flux is present. 
These analyses have been done approximately adopting the
leading order in the powers of $1/r$, but we briefly commented on the
exact condition for the formation of a non-expanding
photon surface and a DTTS near future null infinity for a Vaidya spacetime.
The behavior of null geodesics near null infinity in four and higher dimensions 
differs even when $\dot{m}$ is not so large.
Equivalent properties have been also seen in the extrinsic curvature of an $r$-constant hypersurface.
A similarity between our results and the memory effect can be recognized, since the memory effect\blue{ also gives nontrivial difference between four dimensions and higher dimensions due to asymptotic behavior of the spacetimes}. 
We have also seen   
the contributions from gravitational waves  at the leading order of the Riemann tensor on 
an $r$-constant hypersurface only in four dimensions. 

Finally, we evaluate the order of the energy flux
required for the formation of non-expanding photon surface
or the DTTS near future null infinity. For simplicity, we focus on the spherically
symmetric case, in which the condition of Eq.~\eqref{asy_psf_condition}
is reduced to $\dot{m}(u)=-2$. Using Eq.~\eqref{M(u)}, we find  
$\dot{M}(u)c^2= -c^5/G \sim -4\times10^{59}~\rm{erg/s}$, which is the Planck luminosity \cite{dyson}. Since 
this is conjectured to be the maximum luminosity in the Universe \cite{thorne}, 
its realization near future null infinity would be rather difficult. Therefore, it is expected that the photon 
emitted in the angular directions near future null infinity will reach future null infinity for almost all 
cases in four dimensions. However,
it is \red{mathematically} interesting to explore the possibility
for the construction of some concrete but unfamiliar examples,
and this topic will be addressed in near future. 

%
%

\acknowledgments

M. A. is grateful to Professor S. Mukohyama and Professor T. Tanaka for continuous encouragements and useful suggestions. 
T. S. and K. I. are supported by Grant-Aid for Scientific Research from Ministry of Education, Science, Sports 
and Culture of Japan (Nos. JP17H01091, JP21H05182). T. S., K. I. and H. Y. are also supported by JSPS(No. JP21H05189). 
T. S. is also supported by JSPS Grants-in-Aid for Scientific Research (C) (JP21K03551). K.~I. is also supported by 
JSPS Grants-in-Aid for Scientific Research (B) (JP20H01902).  H.~Y. is in part supported by 
JSPS KAKENHI Grant Numbers JP17H02894 and JP18K03654. The work of H.Y. is partly supported by Osaka City 
University Advanced Mathematical Institute (MEXT Joint Usage/Research Center on Mathematics and Theoretical 
Physics JPMXP0619217849).

\appendix

%
\section{Explicit construction in a Vaidya spacetime}
\label{Appendix:Vaidya}

In this appendix, we perform exact calculations for finding
the condition for an $r$-constant
hypersurface to be a photon surface in a four-dimensional Vaidya spacetime.
The metric of the Vaidya spacetime is 
\begin{equation}
  ds^2 = -f(u,r)du^2-2dudr+r^2\omega_{IJ}dx^Idx^J,
\end{equation}
where $f(u,r)=1-2M(u)/r$ and $M(u)$ is the Bondi mass. We span the standard
spherical-polar coordinates $(\theta,\phi)$ on the unit sphere.
\blue{In this spacetime, the non-zero component of the stress-energy tensor is 
\ba
T_{uu}&=&-\frac{1}{4\pi}\frac{\dot{M}}{r^2},
\ea
\blue{to} which we can apply the fall-off behavior of Eq. \eqref{metric}.}

We study the behavior of a photon in this spacetime.
Due to the spherical symmetry, it is sufficient to study on the
equatorial plane, $\theta = \pi/2$.
Since the system is axially symmetric,
the angular momentum is conserved:
\begin{equation}
L=r^2\phi^\prime.
\end{equation}
The null condition gives
\begin{equation}
-fu^{\prime 2} - 2u^\prime r^\prime + \frac{L^2}{r^2} = 0,
\end{equation}
and the $r$-component of the geodesic equations is
\begin{equation}
  r^{\prime\prime} + \frac12(\partial_uf+f\partial_rf)u^{\prime 2}
  +\partial_rf u^\prime r^\prime - fr\left(\frac{L}{r^2}\right)^2 = 0.
\end{equation}
Combining these equations, we obtain
\begin{equation}
  r^{\prime\prime}
  +\frac{2\dot{M}}{fr} u^\prime r^\prime
  -\frac{L^2}{fr^3}
  \left[\left(1-\frac{2M}{r}\right)\left(1-\frac{3M}{r}\right)+\dot{M}\right]
  \ =\ 0.
  \label{Vaidya_geodesic_eq}
\end{equation}

We now examine the condition for an $r$-constant hypersurface
to be a photon surface. This condition is derived by requiring
$r^\prime = 0$ and $r^{\prime\prime} = 0$. Then, we have
\begin{equation}
\dot{M} = -\left(1-\frac{2M}{r}\right)\left(1-\frac{3M}{r}\right).
  \label{Vaidya_PS_condition}
\end{equation}
Since the Bondi mass $M(u)$ and the quantity $m(u, x^I)$
in the Bondi coordinates are related as $m = 2M$ in four dimensions,
we have the approximate condition of Eq.~\eqref{asy_psf_condition}
by ignoring the terms of $\mathcal{O}(M/r)$.

\blue{So far, we have considered the condition for $S_r$ to be an appropriate photon surface by ignoring higher order terms in the $r^{-1}$ expansion. 
It is also possible for the portion of $S_r$ to be an exact photon surface with $M(u)$ satisfying Eq.~\eqref{Vaidya_PS_condition}.}
\blue{\red{Note that Eq.~\eqref{Vaidya_PS_condition} cannot be satisfied eternally} because the total mass is finite. Here, we consider the finite time interval $0\red{<} u\red{<} u_f$ in which Eq.~\eqref{Vaidya_PS_condition} is exactly satisfied.}
Let $M_0$ be the \red{Bondi mass at $u=0$. Consider the situation in which the Bondi mass decreases intensively so that Eq.~\eqref{Vaidya_PS_condition} is exactly satisfied in the interval $0\red{<} u\red{<} u_f$}\red{.}
The dependence of $M$ on
the $u$ coordinate must be fine tuned as
\begin{equation}
  M(u) = M_0-
  \frac{r\left(e^{u/r}-1\right)\left(1-{2M_0}/{r}\right)\left(1-3M_0/r\right)}
       {3\left(1-2M_0/r\right)-2\left(1-3M_0/r\right)e^{u/r}}.
\end{equation} 
For this expression, the portion of \blue{$0\red{<} u\red{<} u_f$}
of $S_r$ becomes the photon surface\red{. Here, the end of the interval is }\blue{$u_f=r\log[(1-2M_0/r)/(1-3M_0/r)]$}\red{, and then the Bondi mass becomes zero, {\it i.e.} $M(u_f)=0$}.
\blue{After $u=u_f$, $M(u)$ must be zero due to non-negativity of the Bondi mass}.

For a two-dimensional section of a $u$-constant 
\blue{hyper}surface and an $r$-constant hypersurface 
to be a\blue{n exact} DTTS, $r^\prime = 0$ and $r^{\prime\prime}\le 0$
in Eq.~\eqref{Vaidya_geodesic_eq} is the necessary and sufficient
condition.
This leads to the condition of Eq.~\eqref{DTTS_condition_Vaidya}.


\begin{thebibliography}{99}

\bibitem{Abbott:2016blz}
B.~P.~Abbott \textit{et al.} (LIGO Scientific and Virgo Collaborations),
``Observation of Gravitational Waves from a Binary Black Hole Merger,''
Phys. Rev. Lett. \textbf{116}, 061102 (2016).

\bibitem{Akiyama:2019cqa}
K.~Akiyama \textit{et al.} (Event Horizon Telescope Collaboration),
``First M87 Event Horizon Telescope results. I. The shadow of the supermassive black hole,''
Astrophys. J. Lett. \textbf{875}, L1 (2019).



\bibitem{Bondi} 
H.~Bondi, M.~G.~J.~van der Burg and A.~W.~K.~Metzner,
``Gravitational waves in general relativity. VII. Waves from axisymmetric isolated systems,''
Proc. Roy. Soc. Lond. A \textbf{269}, 21-52 (1962).

\bibitem{Sachs} 
R.~K.~Sachs,
``Gravitational waves in general relativity. VIII. Waves in asymptotically flat space-times,''
Proc. Roy. Soc. Lond. A \textbf{270}, 103-126 (1962).

\bibitem{Penrose:1962ij}
R.~Penrose,
``Asymptotic properties of fields and space-times,''
Phys. Rev. Lett. \textbf{10}, 66-68 (1963).

\bibitem{Penrose:1965am}
R.~Penrose,
``Zero rest mass fields including gravitation: Asymptotic behavior,''
Proc. Roy. Soc. Lond. A \textbf{284}, 159 (1965).

\bibitem{Ashtekar:1978}
A.~Ashtekar and R.~O.~Hansen, 
``A unified treatment of null and spatial infinity in general relativity. I. 
Universal structure, asymptotic symmetries, and conserved quantities at spatial infinity,''
J. Math. Phys. \textbf{19}, 1542(1978).

\bibitem{Hollands:2003xp}
S.~Hollands and A.~Ishibashi,
``Asymptotic flatness at null infinity in higher dimensional gravity,''
arXiv:hep-th/0311178.

\bibitem{Hollands:2003ie}
S.~Hollands and A.~Ishibashi,
``Asymptotic flatness and Bondi energy in higher dimensional gravity,''
J. Math. Phys. \textbf{46}, 022503 (2005).


\bibitem{Tanabe:2009va}
K.~Tanabe, N.~Tanahashi and T.~Shiromizu,
``On asymptotic structure at null infinity in five dimensions,''
J. Math. Phys. \textbf{51}, 062502 (2010).

\bibitem{Tanabe:2011es}
K.~Tanabe, S.~Kinoshita and T.~Shiromizu,
``Asymptotic flatness at null infinity in arbitrary dimensions,''
Phys. Rev. D \textbf{84}, 044055 (2011).





\bibitem{no-memory}
S.~Hollands, A.~Ishibashi and R.~M.~Wald,
``BMS Supertranslations and Memory in Four and Higher Dimensions,''
Class. Quant. Grav. \textbf{34}, no.15, 155005 (2017).

\bibitem{Zeldovich:1974}
Y.~B.~Zel'dovich and A.~G.~Polnarev,
``Radiation of gravitational waves by a cluster of superdense stars,''
Sov. Astron. \textbf{18}, 17 (1974)

\bibitem{Christodoulou:1991}
D.~Christodoulou,
``Nonlinear nature of gravitation and gravitational wave experiments,''
Phys. Rev. Lett. \textbf{67}, 1486-1489 (1991).

\bibitem{Thorne:1992}
K.~S.~Thorne,
``Gravitational-wave bursts with memory: The Christodoulou effect,''
Phys. Rev. D \textbf{45}, no.2, 520-524 (1992).

\bibitem{Strominger:2014pwa}
A.~Strominger and A.~Zhiboedov,
``Gravitational Memory, BMS Supertranslations and Soft Theorems,''
JHEP \textbf{01}, 086 (2016). 

\bibitem{Amo:2021gcn}
M.~Amo, K.~Izumi, Y.~Tomikawa, H.~Yoshino and T.~Shiromizu,
``Asymptotic behavior of null geodesics near future null infinity: Significance of gravitational waves,''
Phys. Rev. D \textbf{104}, 064025 (2021).


\bibitem{Cao:2021mwx}
L.~M.~Cao, L.~Y.~Li and L.~B.~Wu,
``Bound on the rate of Bondi mass loss,''
Phys. Rev. D \textbf{104} (2021) no.12, 124017.

\bibitem{Virbhadra:1999nm}
K.~S.~Virbhadra and G.~F.~R.~Ellis,
``Schwarzschild black hole lensing,''
Phys. Rev. D \textbf{62}, 084003 (2000).

\bibitem{Claudel:2000} 
C.~M.~Claudel, K.~S.~Virbhadra, and G.~F.~R.~Ellis,
``The Geometry of photon surfaces,''
J.\ Math.\ Phys.\  {\bf 42}, 818 (2001).

\bibitem{Shiromizu:2017ego}
T.~Shiromizu, Y.~Tomikawa, K.~Izumi and H.~Yoshino,
``Area bound for a surface in a strong gravity region,''
PTEP \textbf{2017}, no.3, 033E01 (2017).

\bibitem{Yoshino:2017gqv}
H.~Yoshino, K.~Izumi, T.~Shiromizu and Y.~Tomikawa,
``Extension of photon surfaces and their area: Static and stationary spacetimes,''
PTEP \textbf{2017}, no.6, 063E01 (2017).

\bibitem{Siino:2019vxh}
M.~Siino,
``Causal concept for black hole shadows,''
Class. Quant. Grav. \textbf{38}, no.2, 025005 (2021).

\bibitem{Yoshino:2019dty}
H.~Yoshino, K.~Izumi, T.~Shiromizu and Y.~Tomikawa,
``Transversely trapping surfaces: Dynamical version,''
PTEP \textbf{2020}, no.2, 023E02 (2020).

\bibitem{Cao:2019vlu}
L.~M.~Cao and Y.~Song,
``Quasi-local photon surfaces in general spherically symmetric spacetimes,''
Eur. Phys. J. C \textbf{81}, 714 (2021).
\blue{\bibitem{Koga:2019uqd}
Y.~Koga and T.~Harada,
``Stability of null orbits on photon spheres and photon surfaces,''
Phys. Rev. D \textbf{100}, no.6, 064040 (2019).}
  
\bibitem{Ishibashi:2007kb}
A.~Ishibashi,
``Higher Dimensional Bondi Energy with a Globally Specified Background Structure,''
Classical Quantum Gravity \textbf{25}, 165004 (2008).

\bibitem{Perlick:2005jn}
V.~Perlick,
``On totally umbilic submanifolds of semi-Riemannian manifolds,''
arXiv:gr-qc/0512066.
\blue{
\bibitem{Geroch}
R. P. Geroch,  ``Asymptotic structure of space-time'', in Asymptotic structure of space-time, ed. Esposite and L. Witten(New York: Plenum, 1977).}

\bibitem{dyson}
F. Dyson, ``Gravitational machines,''
in Interstellar Communication, ed. A.G. Cameron, (New York: Benjamin, 1963), chap 12.

\bibitem{thorne}
K. S. Thorne, ``The Theory of Gravitational Radiation: An Introductory Review,''
in Gravitational Radiation, edited by 
N. Deruelle and T. Piran (North-Holland Company, Amsterdam, New York, Oxford, 1983), p. 1.



\end{thebibliography}
\end{document}